# Magnetization tunable Weyl states in EuB$_6$

Hao Su,[1†] Xianbiao Shi,[2,3†] Jian Yuan,[1†] Xin Zhang,[1] Xia Wang,[1,4] Na Yu,[1,4] Zhiqiang Zou,[1,4] Weiwei Zhao,[2,3] Jianpeng Liu,[1,5*] Yanfeng Guo[1*]

[1] School of Physical Science and Technology, ShanghaiTech University, Shanghai 201210, China

[2] State Key Laboratory of Advanced Welding and Joining, Harbin Institute of Technology, Shenzhen 518055, China

[3] Flexible Printed Electronic Technology Center, Harbin Institute of Technology, Shenzhen 518055, China

[4] Analytical Instrumentation Center, School of Physical Science and Technology, ShanghaiTech University, Shanghai 201210, China

[5] ShanghaiTech Laboratory for Topological Physics, ShanghaiTech University, Shanghai 201210, China

The interplay between magnetism and topological band structure offers extraordinary opportunities to realize rich exotic magnetic topological phases such as axion insulators, Weyl semimetals, and quantum anomalous Hall insulators, which therefore has attracted fast growing research interest. The rare-earth hexaborides EuB$_6$ represents an interesting magnetic topological phase with tunable magnetizations along different crystallographic directions, while the correlation with the topological properties remains scarcely explored. In this work, combining magnetotransport measurements and first principles calculations, we demonstrate that EuB$_6$ exhibits versatile magnetic topological phases along different crystallographic directions, which tightly correlate with the varied magnetizations. Moreover, by virtue of the weak magneto-crystalline anisotropy and the relatively strong coupling between the local magnetization and the conduction electrons, we show that the magnetic ground state of the system can be directly probed by the anisotropy in the magnetotransport properties. Our work thus introduces an excellent platform to study the rich topological phases that are tunable by magnetic orders.

[†]The authors contributed equally to this work.

[*]Corresponding authors:

liujp@shanghaitech.edu.cn,

guoyf@shanghaitech.edu.cn.

**INTRODUCTION**

The correlation between magnetism and nontrivial topological electronic band structure is currently one of the central topics in the field of topological phases of matter. To achieve crucial insights into this issue, topological phases with strong spin-orbit coupling (SOC), low structural symmetry and long-range magnetic order could serve as excellent platform [1]. In such systems, the spin rotation can significantly vary the electronic band structure by the energy of even one order of magnitude larger than the traditional Zeeman splitting, thus allowing for more convenient investigation of the tight link between different spin structures and the accordingly varied topological states. The van der Waals antiferromagnetic (AFM) topological insulator (TI) $MnBi_2Te_4$ is such clear cut example [2-12], on which the application of external magnetic field along specific crystallographic directions can drive the spins to be fully polarized, thus leading to a topological phase transition from AFM TI to ferromagnetic (FM) Weyl semimetal (WSM). Furthermore, the tilting of the Weyl cone could be even controllable in the momentum space via rotating the magnetic field and hence the spin directions [5]. In the Kagome FM WSM $Fe_3Sn_2$, rotating the magnetization directions by external magnetic field could produce varied pairs of Weyl nodes through altering the crystal symmetries [13, 14]. These magnetic topological phases offer more conveniences for the study of correlation between magnetism and topological states, as well as more opportunities for the discovery of intriguing topological properties that could be used in next-generation spintronics. However, such magnetic topological phases are still very rare and the exploration is still extremely desirable.

Recently, several rare-earth hexaboride compounds including $SmB_6$ and $YbB_6$ have been predicted and verified to be correlated TIs or topological Kondo insulators [15-20]. In the meanwhile, another family member of the rare-earth hexaborides, a well-known soft magnetic material $EuB_6$, has also been widely studied due to the novel magnetotransport properties around magnetic phase transition, such as the metal-insulator transition [21, 22], the giant blue shift of the unscreened plasma frequency [23, 24], the large zero-bias anomalies [25] and large negative magnetoresistance [21, 26]. According to previous reports [27, 28], a phase transition from the paramagnetic to FM phase with the moment of $Eu^{2+}$ oriented to [001] direction at ~ 15.3 K was observed. At temperature below ~12.5 K, a new FM phase with moment oriented to [111] direction takes place. Based on the abovementioned two magnetic ground states, theoretical calculations revealed that $EuB_6$ is a topological nodal-line semimetal and a WSM for magnetizations along [001] and [111] directions respectively. Very recently, the angle-resolved photoemission spectroscopy measurements presented evidence for the spontaneous time-reversal symmetry (TRS) broken topological semimetal state in $EuB_6$ [29, 30]. On the other hand, although the above mentioned two magnetic phase transitions have been experimentally observed in $EuB_6$, the specific magnetic structures for them have not been determined yet. Fortunately, the magneto-crystalline anisotropy energy of $EuB_6$ is so small such that the orientation of the magnetization (*M*) can be easily modulated by external magnetic field [31, 32], which may significantly change the topological electronic structure that could be probed by magnetotransport measurements. This is highly reminiscent of the case of a series of soft AFM materials, such as $MnBi_2Te_4/(Bi_2Te_3)_n$ [2-12, 33] $EuCd_2Sb/As_2$ [34-35], GdPtBi [36-38] and $MnSb_2Te_4/(Sb_2Te_3)_n$ [39-47], etc. The weak AFM exchange interactions in these materials can be easily overcome under external magnetic fields that can drive these systems into FM WSMs.

In this work, we demonstrate that $EuB_6$ exhibits versatile magnetic topological phases based on magnetotransport measurements and first principles calculations. Moreover, by virtue of the weak magneto-crystalline anisotropy and the relatively strong coupling between the local *M* and the conduction electrons, we show that the magnetic ground state of the system can be directly probed by the anisotropy in the magnetotransport properties.

The details for crystal growth and basic characterizations, isothermal magnetizations and the magnetotransport data analysis of the $EuB_6$ are presented in the Supplementary Information (SI) which includes Refs. [48 - 54].

The EuB$_6$ crystals were grown by using the similar method described in Ref. [55]. Black crystals with shining surface in a typical dimension of 0.6 ×0.8 ×0.2 mm$^3$ were shown by the picture as an inset of Fig. 1(b). Based on a careful refinement of the single crystal X-ray diffraction patterns, the crystal structure was precisely solved as that shown in inset of Fig. 1(a), in which the Eu atom forms cubic lattice with a B6-octahedron residing inside each unit. Details for the magnetotransport and isothermal magnetizations measurements and first principles calculations could be found in the SI.

**RESULTS AND DISCUSSION**

The resistivity and magnetic properties of EuB$_6$ are depicted in Figs. 1(a) and 1(b), respectively. Seen from the temperature dependence of longitudinal resistivity $\rho_{xx}$ at magnetic field $B = 0$ T with the current $I$ // [100] direction presented in Fig. 1(a), below 100 K, it gradually increases with cooling temperature to ~ 12 K, subsequently exhibits a sudden drop with further decreasing the temperature to 2 K, consistent with the FM phase transition manifested by magnetic susceptibility presented in inset of Fig. 1(b). Therefore, the Kondo-like increase of resistivity should be ascribed to the enhanced scattering of conduction carriers from the Eu magnetic moment due to the critical magnetic fluctuations approaching the ferromagnetic phase, and the formation of magnetic order will then reduce the scattering, leading to the rapid decrease of $\rho_{xx}$. For $B = 0.01$ T, the moment along the [001] direction in ordering state is larger than those along [110] and [111] directions, indicative of strong magnetic anisotropy.

The magnetic field dependence of moments along three directions of $M$ // [001], [110] and [111] directions at 2 K is presented in Fig. 1(b). It should be noted that the moment is zero at $B$ ~ 0 T, which confirms that EuB$_6$ is a soft magnetic material. In previous reports, the simultaneous formation of magnetic domains had been observed by magneto-optical Kerr effect microscopy, implying the intimate link between the topological phase transition and broken time-reversal symmetry [29]. Therefore, we suppose that the reason why the magnetic moment of the magnetic ground state is ~ 0 is that there are opposite directions of spins in different magnetic domains. According to the analysis below, the direction of spin in the magnetic domain tends to along the [001] direction. By increasing $B$, the FM order is apparently enhanced with the saturation moment at 2 K close to 7 $\mu_B$, suggesting that spins of the localized Eu $4f^7$ electrons are actually fully polarized along the magnetic field direction, as shown in Fig. 1(c), and the system eventually enters into the spin-polarized state. The isothermal magnetizations at different temperatures for the three directions are presented in Fig. S2.

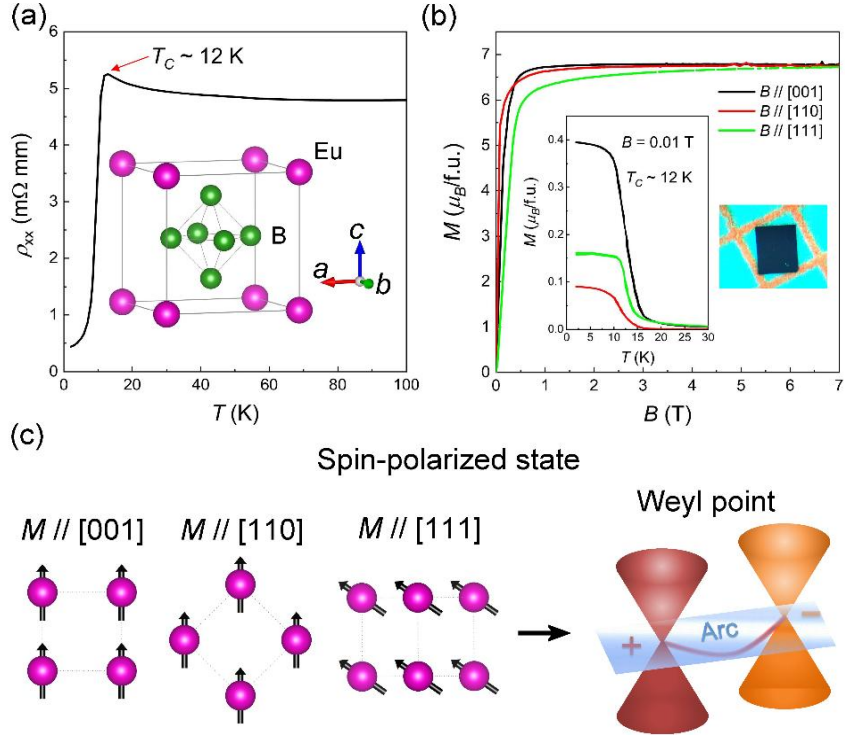

**Fig. 1.** (a) Temperature dependent longitudinal resistivity $\rho_{xx}$. Inset: Schematic crystal structure of EuB$_6$. (b) Isothermal magnetizations measured at 2 K between 0 T – 7 T for $B$ along the [001], [110] and [111] directions, respectively. Inset: Temperature dependent magnetic susceptibility measured at $B$ = 0.01 T and an image of a typical EuB$_6$ single crystal. (c) Schematic spin structures for three different spin-polarized states, which could result in the illustrated Weyl state.

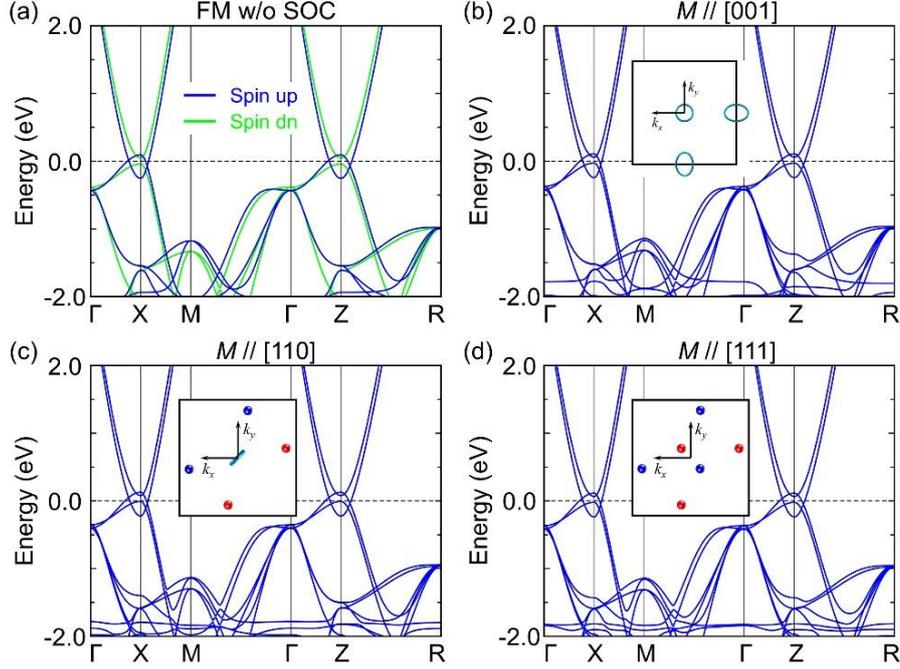

**Fig. 2.** (a) Spin-resolved band structure for the FM state of EuB$_6$ from GGA + U (U = 8 eV) calculations. Band structures calculated using GGA + U + SOC (U = 8 eV) for FM states with the *M* orientation along the (b) [001], (c) [110], and (d) [111] directions. Insets in (b)-(d) illustrate the projections of the nodal-lines (indicated by the cyan lines) configurations and Weyl points (labeled by red/blue dots with different chiralities) distribution on the (001) surface Brillouin zone.

It is necessary to know the electronic band structures corresponding to the different spin-polarized states with *M* along the [001], [110] and [111] directions. As shown in Fig. 2(a), the calculated FM band structure of EuB$_6$ without SOC displays half-semi-metallic behavior. The bands in the spin-down channel exhibit a semiconducting character with a gap of 0.1 eV, whereas the spin-up channel shows a semi-metallic feature with the conduction band being crossed with the valence band. In the presence of SOC, if *M* is along the [001] axis, EuB$_6$ is a Weyl type nodal-line semimetal with two nodal rings centered at X and Y points of the Brillouin zone in the $k_z = 0$ plane and one nodal ring centered at Z point in the $k_z = \pi$ plane, which are protected by the mirror symmetry $M_z$. The nodal lines are schematically plotted in the inset of Fig. 2(b). Once the $M$ orientation is changed from [001] to the [110] axis, the nodal rings in the $M$ // [001] phase will be gapped due to the mirror symmetry breaking, generating two pairs of Weyl points in the $k_z = 0$ plane. In addition to the Weyl points, there is a nodal ring centered at Z point in the vertical diagonal mirror plane as schematically shown in the inset of Fig. 2(c). $EuB_6$ thus is a topological semimetal hosting both Weyl fermions and nodal-line fermions when $M$ is along the [110] direction. Fig. 2(d) shows the electronic structure of $EuB_6$ when $M$ // [111], in which there are three pairs of Weyl nodes. The distributions of these Weyl points in the $M$ // [111] phase are schematically shown in the inset of Fig. 2(d).

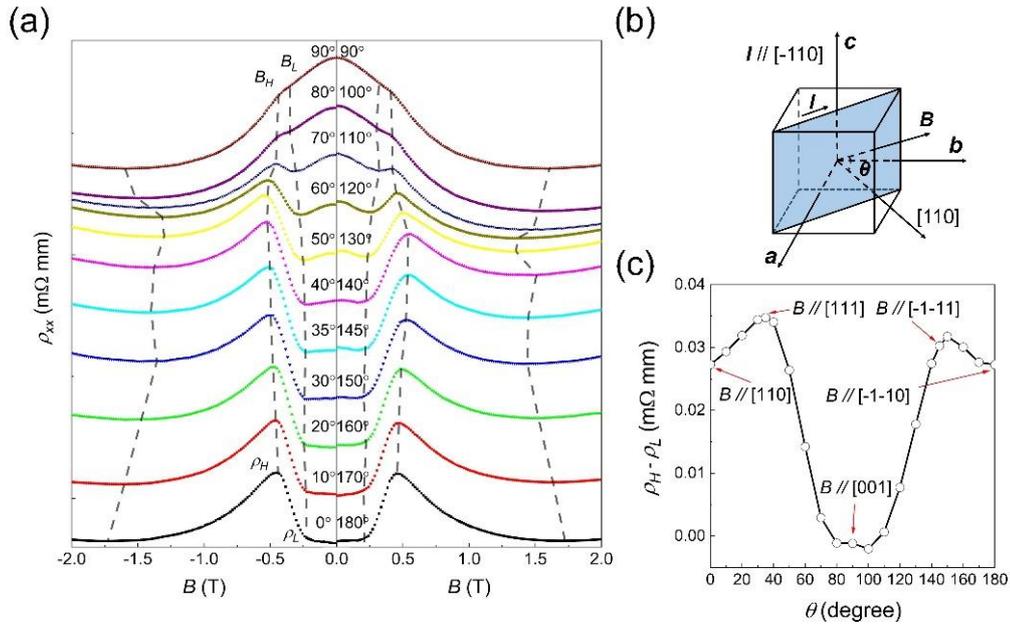

**Fig. 3.** (a) Longitudinal resistivity $\rho_{xx}$ versus magnetic field $B$ at different angles and at the temperature of 2 K. (b) Schematic geometry for the magnetic field rotation measurements. (c) Difference between $\rho_H$ and $\rho_L$ versus angles signifying the scattering between carriers and magnetism during the spin-reorientation transition.

We continue to discuss the magnetotransport properties of $EuB_6$ in different magnetic configurations. The temperature and magnetic field dependent longitudinal resistivities for $M$ along the [001], [110] and [111] directions are presented in Fig. S3.

Fig. 3(a) shows the magnetic field $B$ dependence of longitudinal resistivity $\rho_{xx}$ at different rotation angles at 2 K and Fig. 3(b) shows the geometry for rotation measurements. At $\theta = 0°$, i.e. $B$ // [110] direction, there is a resistivity plateau for weak magnetic field $B < B_L$, where $B_L$ is the threshold magnetic field at which the $\rho_{xx}$ starts to increase, which is commonly observed in some AFM materials with spin-flop transition [33, 40]. By increasing $B$, the scattering between carriers and magnetic moments is enhanced due to the spin-reorientation transition, leading to the increase of resistivity. Thus, the maximum resistivity corresponds to the strongest scattering at $B = B_H$ due to the maximized magnetic fluctuations around the transition point. Here $B_H$ can be considered as a critical magnetic field at which the spin reorientation occurs. When $B > B_H$, the magnetization is re-oriented by the magnetic field, thus the reduced magnetic fluctuations would diminish the resistivity induced by the scattering between electrons and magnetism. The resistivity plateau is preserved until $\theta = 37°$, i.e. $B$ // [111] direction, where the difference between $\rho_H$ and $\rho_L$, i.e., $\rho_H - \rho_L$, reaches the maximum as shown in Fig. 3(c), indicating that the largest scattering-induced variation of magneto-resistivity occurs at this angle. When the angle is further increased, the negative magnetoresistance emerges at $B < B_L$ and it is also accompanied by the decrease of the $\rho_H - \rho_L$. Until the angle reaches 90°, i.e. $B$ //[001] direction, the spin-reorientation induced abrupt enhancement in the magnetoresistance disappears, instead a complete negative magnetoresistance behavior shows up. According to a previous report [29], there are some magnetic domains in EuB$_6$ and the observed intriguing behaviors should be ascribed to the evolution of magnetic domains. By combining the above anisotropic magnetoresistance behaviors, the direction of spin likely tends to be along the [001] direction and other symmetry-equivalent directions. With the $B$ is oriented to [001] direction, the number of magnetic domains in the system is gradually decreased with a gradual increase in the magnetization along [001], resulting in the continuous decrease of resistivity due to the suppressed scattering between electrons and magnetic domain walls. Once the magnetic field is oriented to the [110] or [111] direction, or to a direction deviated from [001], the magnetization along [001] would remain robust for weak magnetic fields due to the magnetocrystalline anisotropy, unless the magnetic field exceeds the threshold field $B_L$, which drives a spin reorientation transition. This explains the emergence of the resistivity plateaus for $B < B_L$, as well as the abrupt enhancement of resistivity when $B \sim B_L$.

It is well known that WSMs with broken TRS, i.e., magnetic WSMs, are also characterized by notable anomalous Hall effect (AHE). The Weyl nodes characterize WSMs can be regarded as "magnetic monopoles" in momentum space, which are the sources generating Berry curvatures in the Brillouin zone. When the Weyl nodes are close to the Fermi level $E_F$, they would contribute to giant net Berry curvature, and generate large intrinsic AHE, which is generally recognized as a fingerprint of the presence of Weyl nodes in a magnetic metallic system. To probe the predicted Weyl state in the FM states of EuB$_6$, the field dependence of the anomalous Hall resistivity $\rho_{xy}^A$ has been extracted and is displayed in Figs. 4(a) - 4(c) for $M$ // [001], [110] and [111], respectively, in the magnetic field range of -9 T to 0 T, and the data which are apparently symmetric with respect to the measured magnetic field directions in the whole magnetic field range from -9 T to 9 T are presented in Fig. S4. Clearly, the EuB$_6$ single crystal exhibits significant AHE at the temperature range of 2 - 20 K, which are obtained by subtracting the nonlinear background fitted by using the two-band model [56], which could be found in more details from Figs. S4-S5 in the SI.

However, in a magnetic topological system, it is necessary to trace the real origin for the AHE since other extrinsic factors such as skew and side-jump contributions besides the Berry curvature could also produce the AHE [57]. To determine the dominant mechanism for the AHE, the so-called TYJ scaling method was used [58], which has been demonstrated to be effective in an array of works. Within the framework TYJ scaling method, the total Hall resistivity could be expressed as $\rho_{xy} = \rho_{xy}^N + \rho_{xy}^A = R_0 B + R_s 4\pi M$, where $R_0$ is the normal Hall coefficient, and $R_s$ is the anomalous Hall coefficient. A more specific formula of $\rho_{xy}^A$ including longitudinal resistivity is $\rho_{xy}^A = a(M)\rho_{xx} + b(M)\rho_{xx}^2$, where the first term denotes the extrinsic contributions including the skew component, while the second term represents the intrinsic contributions also including the side-jump component [59]. Therefore, according to the above formula, $\frac{\rho_{xy}^A}{M\rho_{xx}} = b\rho_{xx}$ is from intrinsic contribution from the spin-polarized states. As shown in Figs. 4(d) - (f), $\rho^A{}_{xy}/(M\rho_{xx})$ are all linearly dependent on $\rho_{xx}$ when $\rho_{xx} < 0.72$ mΩ·mm in three different spin-polarized states, suggesting that the dominant contribution to AHE in EuB$_6$ is from the intrinsic Berry curvatures of the band structures. At 2 K, the converted anomalous Hall conductivity (AHC) $\sigma^A{}_{xy} = -\rho^A{}_{xy} / (\rho^A{}_{xy}{}^2 + \rho_{xx}^2)$ is presented in Figs. 4(g) – 4(i). The $\sigma^A{}_{xy}$ is extracted to be ~ 0.03 mΩ$^{-1}$·mm$^{-1}$ in all spin polarized states, which is rather close to the theoretical value ~ 0.015 - 0.02 mΩ$^{-1}$·mm$^{-1}$ as discussed in Ref.

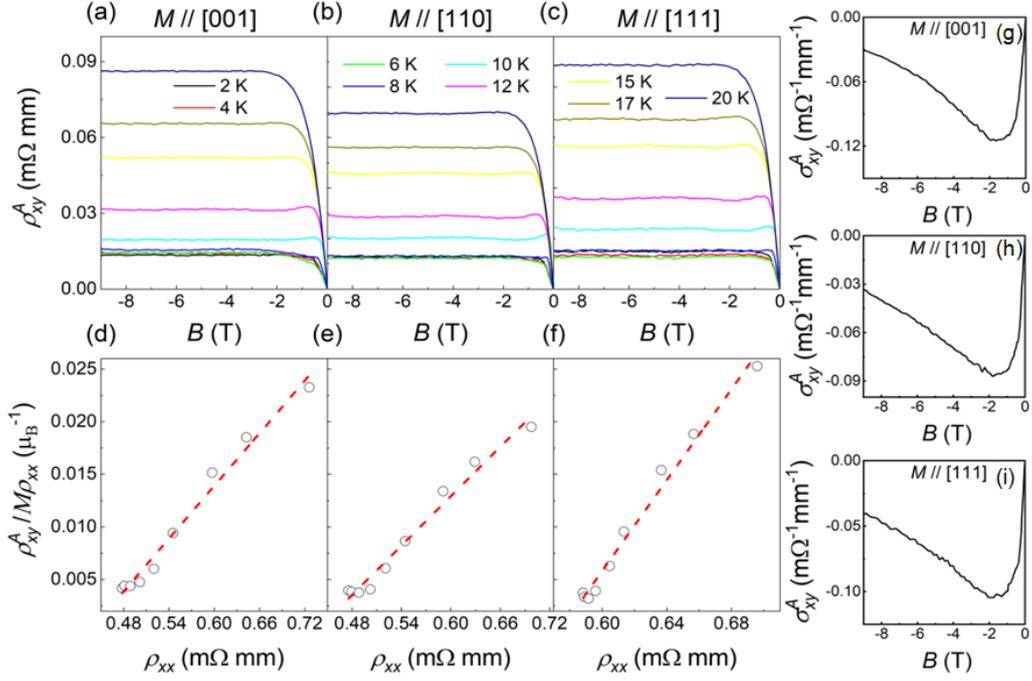

**Fig. 4.** (a) – (c) Anomalous Hall resistivity $\rho^A{}_{xy}$ versus magnetic field at various temperatures. (d) – (f) $\rho^A{}_{xy}/(M\rho_{xx})$ versus $\rho_{xx}$, where the red dashed line denotes the linear fit. (g) – (i) Anomalous Hall conductivity $\sigma^A{}_{xy}$ versus magnetic field at 2 K for $M$ // [001], [110] and [111], respectively.

[31]. On the other side, the side-jump contribution $\sigma^A_{xy,sj}$ can be estimated by the expression $(e^2/(ha))(\varepsilon_{SOC}/E_F)$, where $\varepsilon_{SOC}$ is the SOC energy [60, 61]. Taking the lattice constant $a \sim 4.185$ Å and $\varepsilon_{SOC}/E_F \sim 0.01$ for metallic ferromagnet EuB$_6$, the $|\sigma^A_{xy,sj}|$ was estimated as $9.23 \times 10^{-4}$ mΩ$^{-1}$ mm$^{-1}$, which is almost negligible, thus demonstrating that the AHC is mainly contributed by the Berry curvatures.

**SUMMARY**

In summary, the magneto-crystalline anisotropy and the relatively strong coupling between the local magnetization and the conduction electrons in EuB$_6$ allow for a direct investigation of the topological properties via measuring the magnetotransport properties. With the aid of the first principles calculations, the measurements unveiled that EuB$_6$ actually hosts versatile magnetic topological phases along different crystallographic directions due to the varied magnetizations, thus exposing the imitate correlation between them. The results would be very instructive for the study of magnetic topological physics, in particularly, in such type of topological phases with strong magnetic anisotropy. Moreover, the tunable Weyl states in a single material provide an excellent candidate for use in topological devices with versatile functionalities. According to the theoretical prediction [31], large-Chern-number quantum anomalous Hall effect could be realized in its [111]-oriented quantum-well structures of EuB$_6$. The present study would pave a way toward the realization of the exotic topological properties.

**ACKNOWLEDEMENTS**

The authors acknowledge the support by the National Natural Science Foundation of China (Grant Nos. 92065201, 11874264) and the Shanghai Science and Technology Innovation Action Plan (Grant No. 21JC1402000). J. Liu acknowledges the start-up grant of ShanghaiTech University and the National Key R & D program of China (Grant No. 2020YFA0309601). Y. F. Guo acknowledges the start-up grant of ShanghaiTech University and the Program for Professor of Special Appointment (Shanghai Eastern Scholar). W.W.Z. is supported by the Shenzhen Peacock Team

Plan (Grant No. KQTD20170809110344233) and Bureau of Industry and Information Technology of Shenzhen through the Graphene Manufacturing Innovation Center (Grant No. 201901161514). The authors also thank the support from the Analytical Instrumentation Center (Grant No. SPST-AIC10112914), SPST, ShanghaiTech University.

# Supplementary Information

## Magnetization tunable Weyl states in EuB$_6$

Hao Su,[1†] Xianbiao Shi,[2,3†] Jian Yuan,[1†] Xin Zhang,[1] Xia Wang,[1,4] Na Yu,[1,4] Zhiqiang Zou,[1,4] Weiwei Zhao,[2,3] Jianpeng Liu,[1,5*] Yanfeng Guo[1*]

[1] School of Physical Science and Technology, ShanghaiTech University, Shanghai 201210, China
[2] State Key Laboratory of Advanced Welding and Joining, Harbin Institute of Technology, Shenzhen 518055, China
[3] Flexible Printed Electronic Technology Center, Harbin Institute of Technology, Shenzhen 518055, China
[4] Analytical Instrumentation Center, School of Physical Science and Technology, ShanghaiTech University, Shanghai 201210, China
[5] ShanghaiTech Laboratory for Topological Physics, ShanghaiTech University, Shanghai 201210, China

[†]The authors contributed equally to this work.

[*]Corresponding authors:

liujp@shanghaitech.edu.cn,

guoyf@shanghaitech.edu.cn.

## a. Crystal growth and characterizations

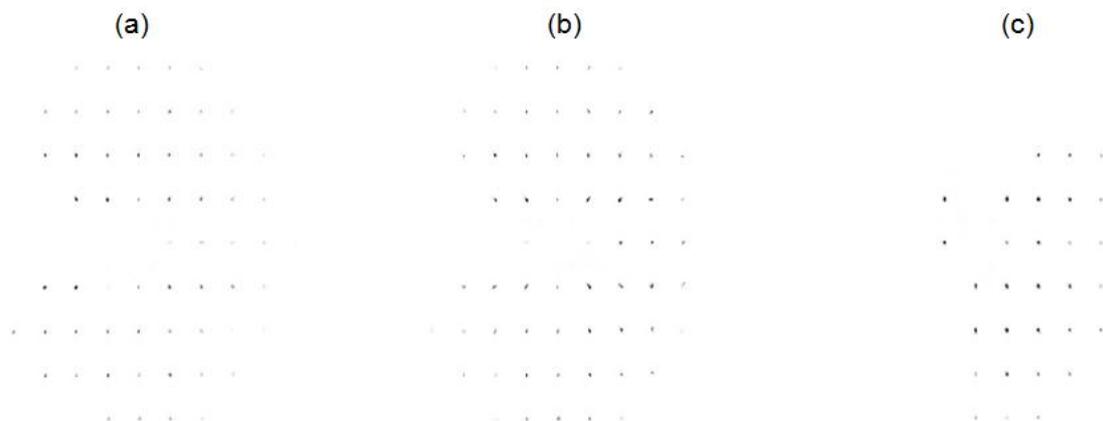

**Fig. S1.** Single crystal X-ray diffraction patterns in the reciprocal space along the (a) (*0kl*), (b) (*h0l*) and (c) (*hk0*) directions for EuB$_6$ measured at 298 K.

EuB$_6$ crystals were synthesized via Al flux method. The starting elements Europium rod (99.9%, Alfa Aesar), Boron powder (99.9%, Macklin), and Aluminum rod (99.9%, Macklin) were taken in the molar ratio 1 : 6 : 120 and sealed into an alumina crucible. All the procedure was done in a glove box filled with argon gas. Then the crucible was put in a high-temperature vacuum atmosphere furnace. The assembly was heated up to 1350 °C, kept for 20 hours at that temperature, then followed by cooling down to 650 °C at a 6 °C/h temperature decreasing rate. The flux was removed by leaching with were separated by washing the crystals by using dilute hydrochloric acid.

The phase and quality examinations of EuB$_6$ were performed on a single-crystal X-ray diffractometer equipped with a Mo Kα radioactive source ($\lambda$ = 0.71073 Å). The diffraction pattern could be satisfyingly indexed on the basis of a CaB$_6$ polytype structure (space group: *cP*7, No. 221) with the lattice parameters $a = b = c = 4.18$ Å, $\alpha = \beta = \gamma = 90°$. These values are very close to the previously reported ones. The clean reciprocal diffraction patterns without other impurity spots indicate the high quality of our single crystals.

## b. Magnetizations and magnetotransport measurements

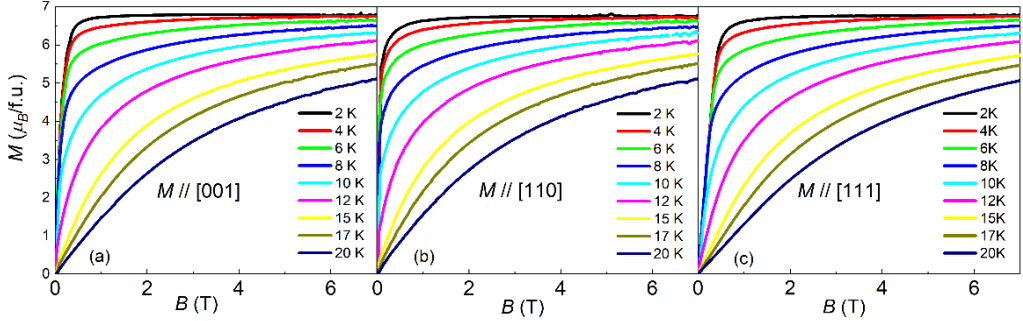

**Fig. S2.** (a)-(c) Isothermal magnetizations at various temperatures for *M* // [001], *M* // [110] and *M* // [111], respectively.

Isothermal magnetizations at various temperatures were measured on a commercial magnetic property measurement system from Quantum Design within the magnetic field range of 0 - 7 T. The results are shown in Fig. S2. Magnetotransport measurements, including the resistivity, magnetoresistance and Hall resistivity, were carried out in a commercial DynaCool Physical Properties Measurement System from Quantum Design. The resistivity and magnetoresistance were measured in a four-probe configuration and the Hall effect measurement was using a standard six-probe method.

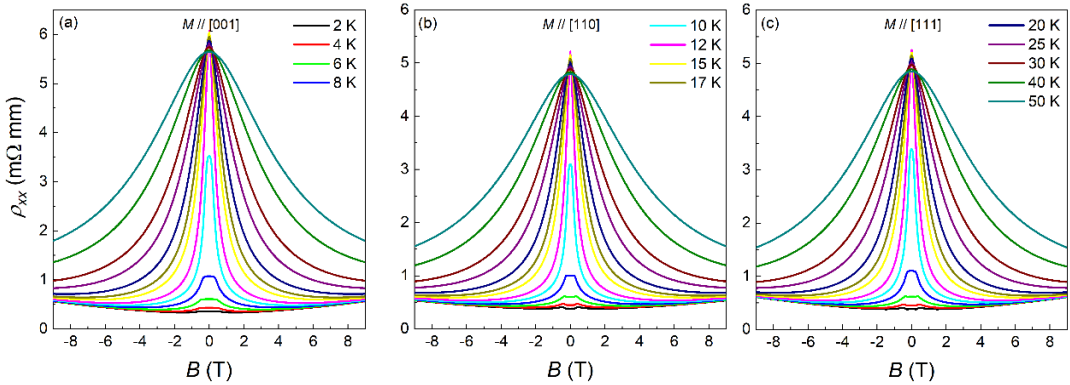

**Fig. S3.** (a)-(c) the longitudinal resistivity at different temperatures with *M* // [001], *M* // [110] and *M* // [111], respectively.

The longitudinal resistivity for *M* // [001], *M* // [110] and *M* // [111] respectively of EuB$_6$ at various temperatures is presented in Fig. S3. Fig. S4 shows the traverse resistivity for *M* // [001], *M* // [110] and *M* // [111] respectively of EuB$_6$ at various temperatures, which are basically nonlinear in the temperature range of 2 K – 20 K. Thus, to expose the carriers for transport, we used the two-band model to fit the data, which is expressed as [48]

$$\sigma_{xy} = \left[\frac{n_h \mu_h^2}{1+(\mu_h B)^2} - \frac{n_e \mu_e^2}{1+(\mu_e B)^2}\right] eB$$

where $n_e$ ($n_h$) denotes the carrier density for the electron (hole), and $\mu_e$ ($\mu_h$) is the mobility of electron (hole), respectively. The fit is fairly nice at the high field part in the spin-polarized state. Figs. S4(d) - 4(f) show the fitting results at 2 K with *M* // [001], *M* // [110] and *M* // [111], respectively. After subtracting the two-band model fitting part indicated in Figs. S4(d) - 4(f), the anomalous Hall resistivity was obtained, which is shown in Fig. S5.

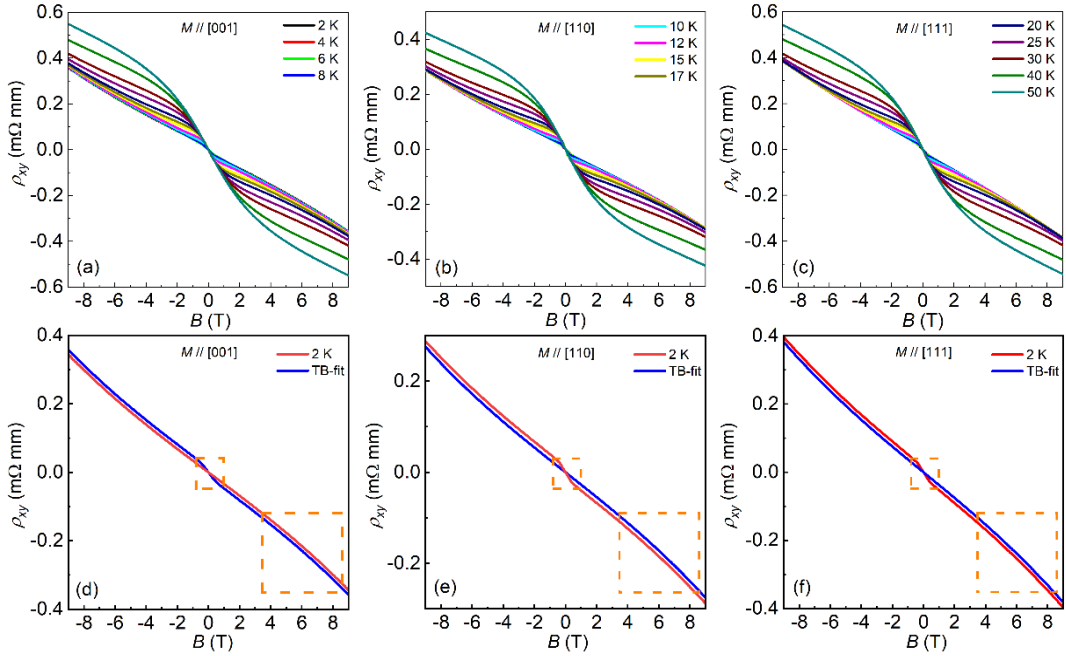

**Fig. S4.** (a)-(c) Transverse resistivity at different temperatures with *M* // [001], *M* // [110] and *M* //

[111], respectively. (d)-(f) The fitting results by using the two-band model at 2 K. The large dash frame zone is the fit part and the small dash frame shows a deviation.

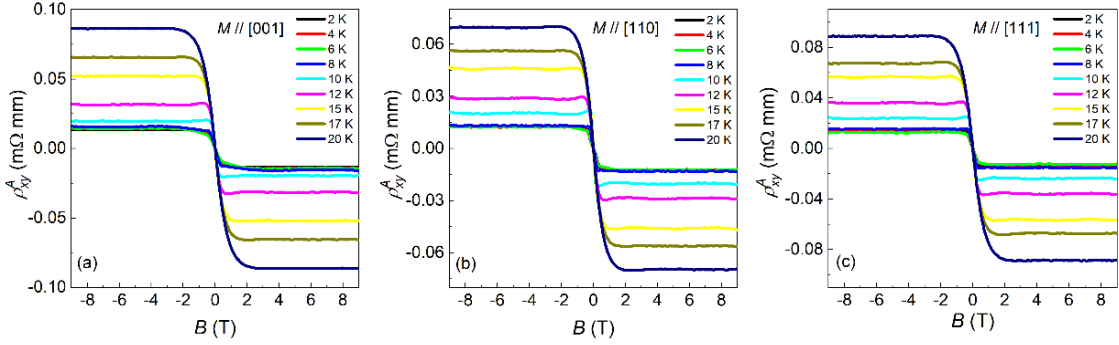

**Fig. S5.** The anomalous Hall resistivity with *M* // [001], *M* // [110] and *M* // [111] in the temperature range of 2 K - 20 K.

## c. First principles calculations

Present first-principles calculations were carried out within the framework of the projector augmented wave (PAW) method [48,49], and employed the generalized gradient approximation (GGA) [50] with Perdew-Burke-Ernzerhof (PBE) [51] formula, as implemented in the Vienna *ab initio* simulation package (VASP) [52-54]. For all calculations, the cutoff energy for the plane-wave basis was set to 500 eV, the Brillouin zone sampling was done with a Γ-centered Monkhorst-Pack k-point mesh of size 12 × 12 × 12, and the total energy difference criterion was defined as $10^{-6}$ eV for self-consistent convergence. The GGA + U scheme was utilized to consider the effect of Coulomb repulsion in the Eu *4f* orbital.